\documentstyle{article}    % Specifies the document style.
\begin{document}           % End of preamble and beginning of text.
\title{Small Radius Perturbation \\ of the Selfgravitating Gas with
Cylindrical Symmetry
\thanks{submited to the Proceeding of Meeting'95 of Istituto per la Ricerca
di Base, Molise, Italy}}
\author{A. Gromov
\thanks{gromov@natus.stud.pu.ru   or   gromov@relat.spb.su}}
\maketitle
\section{Abstract}
Self-consistent mouvement of initial perturbation in density, velocity and
gravitation potentail on the background of the stationary cylindrical
configuration of the gas with gravitation and pressure in Lagrange
variables have been studied. The nonlinear partial differential equation for
description radius motion has been obtained. The linearization of this
equation is reduced to a Klain-Gordon equation which has an analytical
solution.

\section{The Model}
Self-consistent mouvement of initial perturbation in density, velocity and
gravitation potentail on the background of the stationary cylindrical
configuration of the gas with gravitation and pressure in Lagrange
variables have been studied.
In \cite{AD&A} it is shown that a cylindrical symmetry admits an
equilibrium state
of the gas with selfgravitaiting and pressure. The cylinder is supposed
infinity long along the axes and all physical values are dependent on the
radius of cylinder only. The equilibrium is provided by the equality of
gravitational force and force of gas pressure in every point.
In this article the model used in [1] is appeared as a background for a
small perturbations in velocity, density, pressure and gravitataitional
potentail.
In general case the perturbation may depend from two coordinats
and have very complicated geometrical form.
It is not possible to study two-dimensional
nonlinear motion in general case. But the small deviation from the
stationary state admits the principle of superposition. According to this
it is possible to
study radius and longitudinal motions independently and then to sum them
taking into considearation the fact that they are vectors.
This article is dedicated to study pure radius motion.

The model of perturbation depending on radius of cylinder only is choosen.
It means that the cylinder infinite along the axes of symmetry makes
(as a whole) radius motion by the influence of initial perturbation of
stationary state. The perturbation is distributed along the
axes of cylinder thus its value is dependent on the radius coordinat
only. Two causes are able to generate the initial perturbation:
1) the velocity is not equal to zero for particles in the equlibrium state
or 2) displacement of a particle from the equilibrium point is not equal to
zero. The combination of the both is possible as well.

The mathematical description of this model is represen\-ted by
the Cauchy problem for three-dimensional nonstationary partial differential
equations of motion, continuity, Poisson equation
and algebraic equation of state:
%   1
\begin{equation}
\frac{\partial \vec v}{\partial t}+(\vec v \cdot \nabla)
\vec v  =
-\frac{\nabla P}{\rho}-\nabla \Phi
\label{1}
\end{equation}
%   2
\begin{equation}
\frac{\partial \rho}{\partial t}+\nabla \cdot (\rho \vec v) = 0
\label{2}
\end{equation}
%   3
\begin{equation}
{\nabla}^2 \Phi = 4\pi G \rho
\label{3}
\end{equation}
%   4
\begin{equation}
P = A\rho^{\gamma}
\label{4}
\end{equation}
%   5
whis initial conditions
\begin{equation}
\left.\vec v \right|_{t = 0} = \vec v(\vec r,0), \ \ \ \ \
\left.P \right|_{t = 0} = P(\vec r,0), \ \ \ \ \
\left.\Phi \right|_{t = 0} = \Phi(\vec r,0), \ \ \ \ \
\left.\rho \right|_{t = 0} = \rho(\vec r,0)
\label{4A}
\end{equation}
where $P$ is pressure, $\vec v$ - speed, $\Phi$  -
gravitational
potential, $\rho$ - gas density, $t$ - time,
$A=const$, $1 \leq \gamma \leq 2$. $\vec v(\vec r,0)$, $P(\vec r,0)$,
$\Phi(\vec r,0)$ and $\rho(\vec r,0)$ are specified functions.
The system (\ref{1}) - (\ref{4A}) describes hydrodynamic motion in
the ideal classic gas with selfgravitation and pressure.

\section {Characteristic Values and Parameters}

Characteristic values are needed to transfer from dimension to
dimensionless equations. A choise of characteristic values is defindent on
the physical model. Having been choosen these values constitute
dimensionless coefficients of the equations. Depending on the initial
and/or boundary conditions these coefficients are able to play a role of
small or big parameters. There are two kindes of characteristic dimensional
values in this problem. One kind of them is connected with stationary gas
configuration (the
background) and the second one - with perturbations expanded on the
stationary
background. Some various perturbations dependent on the initial
conditions have distinction by different time characteristics (a
period, for example). To compare them it is essentially to have a time
scale not dependent on the perturbation. There is only one choice to
do so - to attach the time scale to the background.

The stationary configuration of the gas will be discribed by a series of
the characteristic values - Jeans lenght $L_0$, gravitaitional potential
$\Phi_0$, gas pressure $P_0$, mass of particle $m$,
concentration of particles $N_0$, density $\rho_0$,
themperature $T_0$, sound speed $c_0$:
%   7
\begin{equation}
L_{0}^{2} = \frac{\pi c_0^2}{G \rho_{0}} \ \ \ \ \ \
\Phi_0 = \frac{G m}{L_{0}} \ \ \ \ \
P_{0} = N_0 k T_0 \ \ \ \ \
{c_0}^2=A{\gamma}{\rho_0}^{\gamma-1}
\label{6}
\end{equation}
that is why:
\begin{equation}
t_{scale} = \frac{1}{\sqrt{G \rho_0}}.
\label{6_6}
\end{equation}
Full enegy of the gas consists of gravitaitional energy and energy of the
gas thermal expansion. The full energy is distribited between these
two components depending on the adiabatic index $\gamma$. The function
$\mu(\gamma)$ dicribes this distribution as follows:
%   14
\begin{equation}
\mu(\gamma) = \frac{k T_{0}}{m \Phi_{0}}.
\label{11}
\end{equation}
Equations (\ref{6}) - (\ref{11}) give the correlation between a number of
particals in the cube with rib length equal to Jeans length and
parameter $\mu(\gamma)$:
%   15
\begin{equation}
N_{0} L_{0}^3 = \pi \gamma \ \mu(\gamma).
\label{13}
\end{equation}
The perturbation will be characterized by the following values definded
by the initial conditions:
the length of wave $\lambda_0$, period of
perturbation $t_0$, velosity $v_0$<
characteristic Mach number and characteristic
dimensionless wave length. They are:
%   13
\begin{equation}
\lambda_0 = v_0 t_0 \ \ \ \ \
M_0 = \frac{v_0}{c_0} \ \ \ \ \ \
\kappa_0 = \frac{\lambda_0}{L_0}.
\label{12}
\end{equation}
A parameter $\alpha$ will also be used:
\begin{equation}
\alpha = \frac{t_0}{t_{scale}}.
\label{T_0}
\end{equation}
 From (\ref{6}), (\ref{12}) and (\ref{T_0}) it follows:
\begin{equation}
\alpha = \frac{\sqrt{\pi} \kappa_0}{M_0}.
\label{15}
\end{equation}

\section{The Initial System of Equations in the Lagrangian Variables}

To convert from dimesionall equation (\ref{1}) - (\ref{4A}) to dimesionless
ones the dimensionless radius, density, pressure, gravitaitional
potentail and velosity are introduced according to the following rules:
%   6
\begin{equation}
\xi=\frac{r}{L_0} \ \ \ \ \
\delta=\frac{\rho}{\rho_0} \ \ \ \ \
p=\frac{P}{P_0} \ \ \ \ \
\varphi=\frac{\Phi}{\Phi_0} \ \ \ \ \
\mbox{v} = \frac{v}{v_0}.
\label{5_5}
\end{equation}
In the cylindrical system of coordinates the dimesionless radius component
of equations (\ref{1}) - (\ref{4A}) are
%   9
\begin{equation}
\frac{\partial \mbox{v}}{\partial \tau} +
a_1 \mbox{v} \frac{\partial \mbox{v}}{\partial \tau}
= - \frac{a_2}{\delta}\frac{\partial P}{\partial \xi} -
a_3 \frac{\partial \varphi}{\partial \xi}
\label{7}
\end{equation}
%   10
\begin{equation}
\frac{\partial \rho}{\partial \tau} + \frac{a_1}{\xi}
\frac{\partial \left(\xi \delta \mbox{v}\right)}{\partial \xi} = 0
\label{8}
\end{equation}
%   11
\begin{equation}
\frac{1}{\xi}\frac{\partial}{\partial \xi}\left(\xi\frac{\partial
\varphi}{\partial \xi}\right) = a_4 \delta
\label{9}
\end{equation}
%   12
\begin{equation}
p =\delta^{\gamma}
\label{4B}
\end{equation}
with initial conditions
\begin{equation}
\left. \mbox {v} \right|_{t = 0} = \mbox{v}(\xi,0) \ \ \ \ \
\left.p \right|_{t = 0} = p(\xi,0) \ \ \ \ \
\left.\phi \right|_{t = 0} = \phi(\xi,0) \ \ \ \ \
\left.\delta \right|_{t = 0} = \delta(\xi,0)
\label{ini_}
\end{equation}
where
$\mbox{v}(\xi,0)$, $p(\xi,0)$, $\phi(\xi,0)$ and $\delta(\xi,0)$  are
specified functions and
\begin{equation}
a_1 = \frac{v_0 t_0}{L_0} \ \ \ \ \
a_2 = \frac{P_0 t_0}{\rho_0 L_0 v_0} \ \ \ \ \
a_3 = \frac{\Phi_0 t_0}{L_0 v_0} \ \ \ \ \
a_4 = \frac{4 \pi G \rho_0 L_0^2}{\Phi_0}.
\label{10}
\end{equation}
According to the (\ref{6}) and (\ref{5_5}) the expressions for $a_1$ -
$a_4$ become:
%   16
\begin{equation}
a_1 = \frac{\alpha M_0}{\sqrt{\pi}} \ \ \ \ \
a_2 = \frac{\alpha M_0}{\gamma \sqrt{\pi}} \ \ \ \ \
a_3 = \frac{\alpha}{
\sqrt{\pi} \gamma \mu(\gamma) M_0} \ \ \ \ \
a_4 = 4 \pi^2 \gamma \mu(\gamma). \nonumber
\label{14}
\end{equation}
Finally in Euler variables the equations being looked for are obtained by
excluding the $\alpha$ from the (\ref{14}) and substituting the result
into the (\ref{7}) - (\ref{9}):
%   18
\begin{equation}
\frac{1}{\kappa_0}\frac{\partial \mbox{v}}{\partial \tau} +
\mbox{v} \frac{\partial \mbox{v}}{\partial \xi}
= - \frac{1}{\gamma \delta}\frac{\partial P}{\partial \xi} -
\frac{1}{\gamma \mu(\gamma) M_0^2}\frac{\partial \varphi}{\partial \xi}
\label{16}
\end{equation}
%   19
\begin{equation}
\frac{\partial \delta}{\partial \tau} + \frac{\kappa_0}{\xi}
\frac{\partial \left(\xi \delta \mbox{v}\right)}{\partial \xi} = 0
\label{17}
\end{equation}
%   20
\begin{equation}
\frac{1}{\xi}\frac{\partial}{\partial \xi}\left(\xi\frac{\partial
\varphi}{\partial \xi}\right) = 4 \pi^2 \gamma \mu(\gamma)\delta
\label{18}
\end{equation}
with initial conditions (\ref{ini_}).
But the problem under consideration may be solved in the Lagrange
coordinates only.
Applying the transformation rules (\cite{S&P}):
\begin{equation}
\frac{\partial}{\partial \tau} \rightarrow \frac{\partial}{\partial \tau} -
\xi \delta \mbox{v} \frac{\partial}{\partial \sigma} \ \ \ \ \ \ \ \
\frac{\partial}{\partial \xi} \rightarrow
\xi\delta\frac{\partial}{\partial \sigma} \ \ \ \ \ \ \ \
\mbox{where} \ \ \ \ \ \ \ \
\sigma = \int \limits_0^\xi \delta \xi d\xi
\label{19}
\end{equation}
($\sigma$ is a dimensionless Lagrange mass variable)
to eqations (\ref{16}) - (\ref{18}) gives the system of equations
describing the problem under consideration in Lagrangian system of
variables:
\begin{equation}
\frac{1}{\kappa_0}
\frac{\partial \mbox{v}}{\partial \tau} =
- \frac{\xi}{\gamma}
\frac{\partial p}{\partial \sigma} -
\frac{\delta \xi}{\gamma\mu(\gamma) M_0^2}\frac{\partial \varphi}{\partial
\sigma}
\label{20}
\end{equation}
\begin{equation}
\frac{\partial \delta}{\partial \tau} +
\kappa_0 \delta^2 \frac{\partial (\mbox{v} \xi)}{\partial \sigma} = 0
\label{21}
\end{equation}
\begin{equation}
\frac{\partial}{\partial \sigma}
\left(\xi^2 \delta \frac{\partial \varphi}{\partial \sigma} \right) =
4 \pi^2\gamma\mu(\gamma)
\label{22}
\end{equation}
with initial conditions:
\begin{eqnarray}
\left. \mbox {v} \right|_{t = 0} = \mbox{v}(\sigma,0) \ \ \ \ \
\left.p \right|_{t = 0} = p(\sigma,0) \ \ \ \ \
\left.\phi \right|_{t = 0} = \phi(\sigma,0) \ \ \ \ \
\left.\delta \right|_{t = 0} = \delta(\sigma,0) \nonumber\\
\left.\xi \right|_{t = 0} = \xi(\sigma,0) \ \ \ \ \
\delta(0,0) = 1
\label{ini_1}
\end{eqnarray}
where
$\mbox{v}(\sigma,0)$, $p(\sigma,0)$, $\phi(\sigma,0)$, $\delta(\sigma,0)$
and $\xi(\sigma,0)$ are specified functions.
The dependence of function $\xi(\sigma,0)$ from $\sigma$ will not be used in
this article and, that is why, will not be studied.

\section{The Development of the Radius Motion Equation}

In this section the system of equations (\ref{20}) - (\ref{22}) will be
reduced to one equation for Euler coordinate $\xi(\sigma,\tau)$.
The dependence on $(\sigma,\tau)$ will be omitted expect the
case $\tau = 0$. It means that $\delta$ means $\delta(\sigma,\tau)$, but for
$\tau = 0$ we will write $\delta(\sigma,0)$.
The Poison equation (\ref{22}) may be integrated directly
with initial condition:
\begin{equation}
\left.\frac{\partial \varphi}{\partial \sigma}\right|_{\sigma=0} = const.
\label{23}
\end{equation}
%        2
Because of the equation (\ref{22}) structure $const$ should be
finite but it is not essential and it doesn't matter which the $const$ is.
The first integral of Poisson equation is:
\begin{equation}
\xi^2 \delta \frac{\partial \phi}{\partial \sigma} =
4\pi\gamma\mu(\gamma)\sigma - L(\tau)
\label{24}
\end{equation}
where $L(\tau)$ is some function will be defined later from the
conditions of independence all physical values from the axes of symmetry.
Excluding the $\xi \delta \frac{\partial \varphi}{\partial \sigma}$ from
(\ref{22}) and (\ref{24}) the equation:
\begin{equation}
\frac{\xi}{\kappa_0}\frac{\partial \mbox{v}}{\partial \tau} =
- \frac{\xi^2}{\gamma}\frac{\partial P}{\partial \sigma}
- \frac{4\pi^2\gamma\mu(\gamma)\sigma - L(\tau)}{\gamma\mu(\gamma)M_0^2}
\label{25}
\end{equation}
is obtained.

The following transformation is connected with the equation of
continuity (\ref{21}). Using the definition of velosity in the Lagrangian
variables:
\begin{equation}
\mbox{v} = \frac{\partial \xi}{\partial \tau}
\label{26}
\end{equation}
and dividing (\ref{21}) by $\delta^2$ the equation:
\begin{equation}
\frac{\partial}{\partial\tau}\frac{1}{\delta} = \frac{1}{\kappa_0}
\frac{\partial^2 \xi^2}{\partial\tau\partial\sigma}
\label{27}
\end{equation}
is obtained. Its integral is:
\begin{equation}
\frac{1}{\delta} - \frac{\kappa_0}{2}\frac{\partial \xi^2}{\partial \sigma} =
G(\sigma) +
%\left(
\frac{1}{\delta(\sigma,0)} -
\frac{\kappa_0}{2}\frac{\partial\xi^2}{\partial\sigma}\left.\right|_{\tau=0}
%\right)
\label{28}
\end{equation}
The function $G(\sigma)$ will be defined from the initial conditions.
 From the definition of $\sigma$ (\ref{19}) and equality:
\begin{equation}
\frac{\partial \xi^2}{\partial \sigma}\left.\right|_{\tau=0} =
\frac{d\xi^2}{d\sigma}
\label{29}
\end{equation}
it follows that:
\begin{equation}
\frac{\partial\xi^2}{\partial \sigma}\left.\right|_{\tau=0} =
\frac{2}{\delta(\sigma,0)}.
\label{30}
\end{equation}
Due to this
\begin{equation}
%\left(
\frac{1}{\delta(\sigma,0)} -
\frac{\kappa_0}{2}\frac{\partial\xi^2}{\partial\sigma}\left.\right|_{\tau=0}
%\right)
=
\frac{1 - \kappa_0}{\delta(\sigma,0)}.
\label{31}
\end{equation}
Finally, the integral of continuity equation is:
\begin{equation}
\frac{1}{\delta(\sigma,0)} = \frac{\kappa_0}{2}\frac{\partial\xi^2}{\partial
\sigma} +
\frac{1 - \kappa_0}{\delta(\sigma,0)} +G(\sigma)
\label{32}
\end{equation}
Let's designate
\begin{equation}
\Sigma(\sigma) =
\frac{1 - \kappa_0}{\delta(\sigma,0)} +G(\sigma).
\label{324}
\end{equation}

The next step is substitution the $\rho$ from (\ref{32}) into the equation
of state (\ref{4B}) and motion equation (\ref{20}):
\begin{equation}
\frac{\xi}{\kappa_0}\frac{\partial^2\xi}{\partial\tau^2} =
- \frac{\xi^2}{\gamma}\frac{\partial}{\partial \sigma}
%%\frac{1}{
\left(
\frac{\kappa_0}{2}\frac{\partial\xi^2}{\partial\sigma}
+
% \frac{1 - \kappa_0}{\delta(\sigma,0)} + G(\sigma)
\Sigma(\sigma)
\right)^{-\gamma}
%%}
- \frac{4\pi^2\gamma\mu(\gamma)\sigma - L(\tau)}{\gamma\mu(\gamma)M_0^2}
\label{33}
\end{equation}
The function $L(\tau)$ is found out from the follows condition:
all physical values are not dependent on the axes of sylinder. Let's use it
for $\sigma = \xi = 0$. It is follws from (\ref{33}):
\begin{equation}
L(\tau) = 0
\label{34}
\end{equation}

The function $G(\sigma)$ is found out from the initial conditions
(\ref{ini_1}).
To calculate it we introduce a parameter $\ddot \xi_0(\sigma)$:
\begin{equation}
\left.\frac{\partial^2\xi}{\partial\tau^2}\right|_{(\sigma,0)} =
\ddot \xi_0(\sigma)
\label{35}
\end{equation}
Using (\ref{30}) - (\ref{324}) and calculating (\ref{33}) for $\tau = 0$
the equation for function $G(\sigma)$ is obtained:
\begin{equation}
G(\sigma)^{'} -
\frac{1}{\gamma}\left(G(\sigma) + \frac{1}{\delta}\right)^{\gamma+1}
\left(\frac{\xi(\sigma,0)\ddot\xi_0}{\kappa_0} -
\frac{4\pi^2\sigma}{M_0^2}\right) - \frac{\delta^{'}}{\delta^2} = 0
\ \ \ \ \ \mbox{where} \ \ \ \ \ {'} = \frac{d}{\sigma}.
\label{36}
\end{equation}
To find out the initial condition for this equation
we calculate the (\ref{32}) at point $\sigma = 0$ and obtain:
\begin{equation}
G(0) = 0.
\label{37}
\end{equation}
So, the equation of radius motion which should be obtained in this section
is:
\begin{equation}
\frac{\xi}{\kappa_0}\frac{\partial^2\xi}{\partial\tau^2} =
- \frac{\xi^2}{\gamma}\frac{\partial}{\partial \sigma}
%\frac{1}{
\left(
\frac{\kappa_0}{2}\frac{\partial\xi^2}{\partial\sigma}
+ \Sigma(\sigma)
%\frac{1 - \kappa_0}{\delta(\sigma,0)} + G(\sigma)
%
\right)^{-\gamma}
%}
- \frac{4\pi^2\sigma}{M_0^2}
\label{38}
\end{equation}
with initial conditions
\begin{equation}
\left.\xi \right|_{t = 0} = \xi(\sigma,0), \ \ \ \ \ \ \ \
\left.\frac{\partial\xi}{\partial \sigma} \right|_{t = 0} =
\dot\xi(\sigma,0)
\label{39}
\end{equation}
where
$\mbox{v}(\sigma,0)$, $p(\sigma,0)$, $\phi(\sigma,0)$, $\delta(\sigma,0)$
and $\xi(\sigma,0)$ are specified functions.
Function $G(\sigma)$ is definded by the equation (\ref{36}) with
initial condition (\ref{37}). The equation (\ref{39}) is nonlinear
nonstationary partial differential equation which cannot be solved
analytically in general but only after linierization.

\section{The Linearization of Radiuse Motion Equation}

The sections $5$ and $6$ ar edevoted to studying the small perturbations of
the
gas. An arbitrary radius motions are described by the equation (\ref{38})
with initial conditions (\ref{39}). According to the Model the
transformation
from (\ref{38}) - (\ref{39}) to linear approximation will be made now.
Let's assume that a particle is in an equilibrium in the point of radius
with
Lagrange coordinat $\sigma_0$. Because of smallness of deviation the
particle from the point of equlibrium the Euler coordinat of the eparticle
may be presented as a sum:
\begin{equation}
\xi(\sigma,\tau) = \xi_0 + \psi(\sigma,\tau) \ \ \ \ \
\mbox{where} \ \ \ \ \ \ \ \
\left|\psi(\sigma,\tau)\right| \ll \xi_0 \ \ \ \ \
\mbox{and} \ \ \ \ \ \xi_0 = \xi(\sigma_0,0)
\label{41}
\end{equation}
To study a generaly case let assume that
the particle is displaced
at the moment of time $\tau = 0$
from the poin of
equilibrium to new poin with lagrange coordinat $\sigma = \sigma_0 +
\triangle \sigma$ and has a velosity equal to $\mbox{v}(\sigma_0 +
\triangle \sigma)$.
So, the initial conditions according this model are
\begin{equation}
\xi(\sigma,0) = \xi(\sigma_0,0) + \psi(\sigma,0)
 \ \ \ \ \ \ \ \
\left.\frac{\partial\xi(\sigma,\tau)}{\partial \sigma}\right|_{\tau=0} =
\dot\psi(\sigma,0)
\label{A41a}
\end{equation}
where
\begin{equation}
\psi(\sigma,0) = \Psi
\delta[\sigma - (\sigma_0 + \triangle \sigma)]  \ \ \ \ \ \ \
%\left.\frac{\partial\psi(\sigma,\tau)}{\partial \sigma}\right|_{\tau=0} =
\dot\psi(\sigma,0) = \dot\Psi\delta[\sigma - (\sigma_0 + \triangle \sigma)]
\label{A41}
\end{equation}
where $\Psi$ and $ \dot\Psi$ are parameters of the problem, and
$\delta[\sigma  - (\sigma_0 + \triangle \sigma)]$ is a delta-fuction.
Let's designate:
\begin{equation}
\bar\sigma = \sigma + \triangle\sigma
\label{040}
\end{equation}
In addition to (\ref{41}) the condition
\begin{equation}
\left|\frac{\kappa_0}{2}\frac{\partial\xi^2}{\partial\sigma}\right| \ll
\Sigma(\sigma)
\label{40}
\end{equation}
will be used for linearisation of the equation (\ref{38}).
(\ref{40}) allows to simplify
\begin{equation}
\left(\frac{\kappa_0}{2}\frac{\partial\xi2^2}{\partial \sigma} +
\Sigma(\sigma)\right)^{\gamma+1} \approx
\Sigma(\sigma)^{\gamma+1} + (\gamma +
1)\frac{\kappa_0}{2}\frac{\partial\xi^2}{\partial\sigma}\Sigma^{\gamma}(\sigma)
\label{42}
\end{equation}
and after substitution the (\ref{42}) into (\ref{38}) the follwing equation
is obained:
\begin{equation}
\left(\frac{1}{\kappa_0\xi}\frac{\partial^2\xi}{\partial\tau^2} +
\frac{4\pi^2}{M_0^2}\frac{\sigma}{\xi^2}\right)
\left(
\Sigma^{\gamma+1} + \kappa_0\frac{\gamma+1}{2}\Sigma^{\gamma}
\frac{\partial\xi^2}{\partial\sigma}\right) =
\frac{\kappa_0}{2}\frac{\partial^2\xi^2}{\partial\sigma^2} + \Sigma^{'}
\label{f}
\end{equation}
where $\Sigma^{'} = \frac{d\Sigma}{d\sigma}$.
To linearized (\ref{f}) use the (\ref{41}). This substitution gives
the equaton
\begin{equation}
\frac{\partial^2\psi}{\partial\tau^2} -
\frac{\kappa_0^2\xi^2_0}{\Sigma^{\gamma+1}_0}\frac{\partial^2\psi}{\partial
\sigma^2} +
\frac{4\pi^2\kappa^2_09\gamma+1)\bar\sigma}{\Sigma_0 M_0^2}
\frac{\partial\psi}{\partial\sigma} +
\frac{4\pi^2\kappa_0\bar\sigma}{M^2_0\xi_0} -
\frac{\kappa_0\xi_0\Sigma^{'}_0}{\Sigma^{\gamma+1}_0} = 0
\label{af}
\end{equation}
To simplify this equation denote
\begin{equation}
w_0^2 = \frac{\kappa_0^2\xi_0^2}{\Sigma^{\gamma+1}_0} \ \ \ \ \
B = \frac{4\pi^2(\gamma+1)\kappa_0^2\bar\sigma}{\Sigma_0M_0^2} \ \ \ \ \
C = \kappa_0\xi_0\frac{\Sigma^{'}_0}{\Sigma^{\gamma+1}_0} -
\frac{4\pi^2\kappa_0\bar\sigma}{M_0^2\xi_0}
\label{44}
\end{equation}
Then the equation (\ref{af}) become as follows:
\begin{equation}
\frac{\partial^2\psi}{\partial\tau^2} -
w_0^2\frac{\partial^2\psi}{\partial\sigma^2} +
B\frac{\partial\psi}{\partial\sigma} - C = 0
\label{43}
\end{equation}
with the initial conditions (\ref{A41}).
%%
%\begin{equation}
%\psi(\sigma,0) = \Psi \ \ \ \ \ \ \ \
%\left.\frac{\partial\psi(\sigma,\tau)}{\partial\tau)}\right|_{\tau=0} =
%\dot\Psi
%\label{t}
%\end{equation}
%%
$B$ has a sense of acceleration. According to this two new characteristic
values are introduced:
\begin{equation}
\sigma^{*} = \frac{w^2_0}{B} \ \ \ \ \ \mbox{and}
 \ \ \ \ \ \tau^{*} = \frac{\sigma^{*}}{w_0}
\label{45}
\end{equation}
The equation (\ref{43}) will be transformated now by the substitution
\begin{equation}
\psi(\sigma,\tau) = u(\sigma,\tau)\exp{\frac{\sigma -
\bar\sigma}{2\sigma^{*}}} + C\left(\tau^{*}\right)^2\frac{\sigma -
\bar\sigma}{\sigma^{*}}
\label{46}
\end{equation}
where $u(\sigma,\tau)$ is a new function. The result of the substitution
is:
\begin{equation}
\frac{\partial^2u}{\partial\tau^2} -
w_0^2\frac{\partial^2u}{\partial\sigma^2} +
\frac{u}{\left(2\sigma^{*}\right)^2} = 0
\label{47}
\end{equation}
For this equation the Cauche problem is calculated with the following
initial conditions which are obtained by transformation according to
(\ref{46}):
begin:
\begin{equation}
u(\sigma,0) = \psi(\sigma,0)
%\delta(\sigma - \sigma_0)
 \ \ \ \ \ \ \ \
\left.\frac{\partial u(\sigma,\tau}{\partial \tau}\right|_{\tau = 0} =
\dot \psi(\sigma,0)
%\delta(\sigma - \sigma_0)
\label{48}
\end{equation}
It is the Klaine-Gordon equation (\cite{Z})

\section{The Solution of the Klaine-Gordon Equation}

To solve the Kline-Gordon equation the Riemann method of the Cauchy problem
solution will be used (\cite{Gl}). The equation (\ref{47}) is a hyperbolic
equation. In the canonic coordinates $(\nu,\eta)$:
\begin{equation}
\nu = \frac{\sigma + w_0 \tau}{2\sigma^{*}w_0} \ \ \ \ \ \ \ \
\eta = \frac{\sigma - w_0 \tau}{2\sigma^{*}w_0}
\label{101}
\end{equation}
the equation (\ref{47}) is reduced to the form:
\begin{equation}
\frac{\partial^2u}{\partial\nu\partial\eta} -
\frac{u}{4} = 0
\label{102}
\end{equation}
with initial conditions (\ref{48}) transformated according to the rules:
(\ref{101}):
\begin{equation}
\left.u\left(\sigma^{*}w_0(\nu +\eta)\right)\right|_{\nu=\eta} = u(\sigma,0)
 \ \ \ \ \ \ \ \
\left.
\frac{1}{2\sigma^{*}}\left(\frac{\partial u}{\partial \nu} -
\frac{\partial u}{\partial \eta}\right)
\right|_{\nu=\eta} =
\left.\frac{\partial u(\sigma,\tau)}{\partial\tau}\right|_{\tau = 0}
\label{103}
\end{equation}
In the Riemann method the solution of the equation (\ref{102}) is
determined by the (\ref{103}) and Riemann function $V(\nu,\eta)$ satisfying
the conjugatee equation:
\begin{equation}
\frac{\partial^2V}{\partial\nu\partial\eta} -
\frac{V}{4} = 0
\label{104}
\end{equation}
with initial conditions:
\begin{equation}
V(\nu,\hat\eta) = 1 \ \ \ \ \ \ \ \
V(\hat\nu,\eta) = 1.
\label{105}
\end{equation}
Let's find out the Riemann function $V$ in the form
\begin{equation}
V = N(p) \ \ \ \ \ \ \ \ \mbox{where} \ \ \ \ \ \ \ \
p = \sqrt{(\nu - \hat\nu)(\hat\eta - \eta)}
\label{106}
\end{equation}
wherer $(\nu,\mu)$ is the coordinate of the point where the solution
found out and $(\hat\nu,\hat\eta)$ belong to the line in which the initial
conditions are given.
 From (\ref{106}) it follows that $\tau = 0$ correspond to $\hat\nu =
\hat\eta$.
The substitution (\ref{106}) into the (\ref{104}) gives the ordinary
differential equation:
\begin{equation}
N^{''} + \frac{1}{p}N^{'} + N(p) = 0
\label{106}
\end{equation}
which has a finite solution:
\begin{equation}
N(p) = J_0\left(\sqrt{(\nu - \hat\nu)(\hat\eta - \eta)}\right)
\label{107}
\end{equation}
where $J_0$
is a Bessel function of null order.
The initial conditions (\ref{105}) are satisfied.
Let's return to the equation (\ref{102}).
According to the Riemann method
the solution of the equation (\ref{102}) satisfying the conditions
(\ref{103}) has a form:
\begin{eqnarray}
u(\nu,\eta)=
\frac{1}{2}
\left\{
u\left[\left.\sigma^{*}w_0(\nu + \eta)\right|_{\nu = \eta}\right]
+
u\left[\left.\sigma^{*}w_0(\nu + \eta)\right|_{\eta = \nu}\right]
\right\} +
 \nonumber \\
\frac{1}{2}\int_{\eta}^{\nu}
J_0\left(\sqrt{(\nu - \hat\nu)(\hat\nu - \eta_0)}\right)
\left.\left(
\frac{\partial u}{\partial \nu} - \frac{\partial u}{\partial \eta}
\right)\right|_{\hat\nu=\hat\eta}
d\hat\nu
- \nonumber \\
\frac{\nu - \eta}{4}
\int_{\eta}^{\nu}
\frac{J^{'}_0
\left(\sqrt{(\nu - \hat\nu)(\hat\nu - \eta)}\right)}
{\sqrt{(\nu -\hat\nu)(\hat\nu - \eta)}}
u\left[\left.\sigma^{*}w_0(\hat\nu + \hat\eta)\right|_{\hat\nu=\hat\eta}\right]
d\hat\nu
\label{108}
\end{eqnarray}
Let's transform now the initial conditions (\ref{A41a}) from the dependence
on $\sigma$ to dependence on $\hat\nu$:
\begin{eqnarray}
u\left[
\left.\sigma^{*}w_0(\nu + \eta)\right|_{\nu=\eta}
\right]
 =
u(\sigma,0) =
\Psi\delta\left(\sigma - \sigma(0)\right) = \nonumber\\
\Psi\delta\left[
\left.\sigma^{*}w_0(\hat\nu + \hat\eta)\right|_{\hat\nu=\hat\eta}
- \sigma(0)
\right] \nonumber\\
\frac{1}{2\sigma^{*}}\left(
\left.
\frac{\partial u}{\partial \nu} - \frac{\partial u}{\partial \eta}
\right|_{\nu=\eta}
\right) =
\left.\frac{\partial u(\sigma,\tau)}{\partial\tau}\right|_{\tau=0} =
\dot\Psi\delta\left(\sigma - \sigma(0)\right) = \nonumber\\
\dot\Psi\delta\left[\left.\sigma^{*}w_0(\hat\nu +
\hat\eta)\right|_{\hat\nu=\hat\eta} -
\sigma(0)\right]
\label{179}
\end{eqnarray}
After substitution the (\ref{179}) and
\begin{eqnarray}
\nu - \eta = \frac{\tau}{\sigma^{*}} \ \ \ \ \ \ \ \
(\nu - \hat\nu)(\hat\nu - \eta) = (w_0\tau)^2 - (\sigma - \sigma(0)^2)
\label{109}
\end{eqnarray}
into (\ref{108}) the solution of the equation (\ref{102}) with initial
conditions (\ref{48}) obtain:
\begin{eqnarray}
u(\sigma,\tau) =
\frac{1}{2}
\left\{
       \delta\left[w_0\tau + (\sigma - \bar\sigma)\right] +
       \delta\left[w_0\tau - (\sigma - \bar\sigma)\right]
\right\}
+ \nonumber\\
\frac{\dot\Psi\sigma(0)}{2w_0}
J_0
\left(
\frac{
      \sqrt{(w_0\tau)^2 - \left(\sigma - \sigma(0)\right)^2}
     }
     {
      2\sigma^{*}w_0
     }
\right)
-
\frac{\tau\Psi w_0}{4}
\frac{
      J_1
\left(
        \frac{
              \sqrt{(w_0\tau)^2 - \left(\sigma - \sigma(0)\right)^2}
             }
             {
              2\sigma^{*}w_0
             }
\right)
     }
     {
       \sqrt{(w_0\tau)^2 - \left(\sigma - \sigma(0)\right)^2}
     }
\label{178}
\end{eqnarray}
This formula together with (\ref{46}) define the solution of linear problem.

\section{Results and Discation}

The obtained solution has a formal character of oscillations.But there are
two reasons due to which this conclusion should be triated critically.
First: This is a solution of a linear equation. The area of the initial
conditions where linear solution differs slightly from the nonlinear one is
not discuss in this article. The second: The solution has been obtained in
the Lagrangian variables but when we are speaking about oscillation Euler
variables are meant. The transformation from Lagrange variables to Euler
sones is not performed in this article also.

\section{Acknowledgements}
I'm grateful to Prof. Arthur D. Chernin for encouragement and discasion.
This paper was financial supported by "COSMION" Ltd., Moskow.
%

%\section{References}

\end{document}